\documentclass[a4paper]{PoS}

\usepackage{amsmath,amssymb} 
\usepackage{caption}
\usepackage{subcaption}
\usepackage{braket}
\usepackage{slashed}
\usepackage{bbold}
\usepackage{graphicx}
\usepackage{fontenc}
\usepackage{times}
\usepackage{mathptmx}
\usepackage[utf8]{inputenc}

\title{Disconnected hadronic vacuum polarization contribution to the muon g-2 with HISQ}

\ShortTitle{Disconnected HVP to the muon g-2}

\author{\speaker{Shuhei Yamamoto}\\
        Department of Physics and Astronomy, University of Utah, Salt Lake City, UT 84112, USA\\
        E-mail: \email{sy3394@physics.utah.edu}}
\author{Carleton DeTar\\
        Department of Physics and Astronomy, University of Utah, Salt Lake City, UT 84112, USA\\
        E-mail: \email{detar@physics.utah.edu}}
\author{Aida X. El-Khadra\\
        Department of Physics, University of Illinois Urbana-Champaign, Urbana, IL 61801-3003, USA \& Fermilab, Batavia, IL 60510-5011 USA\\
        E-mail: \email{axk@illinois.edu}}
\author{Craig McNeile\\
        Centre for Mathematical Sciences, University of Plymouth, Plymouth PL4 8AA, United Kingdom\\
        E-mail: \email{craig.mcneile@plymouth.ac.uk}}
\author{Ruth S. Van de Water\\
        Fermilab, Batavia, IL 60510-5011 USA\\
        E-mail: \email{ruthv@fnal.gov}}
\author{Alejandro Vaquero\\
        Department of Physics and Astronomy, University of Utah, Salt Lake City, UT 84112, USA\\
        E-mail: \email{alexvaq@physics.utah.edu}}   
\author{Fermilab Lattice, HPQCD, and MILC Collaborations}  
        
\abstract{We describe a computation of the contribution to the anomalous magnetic moment of the muon from the disconnected part of the hadronic vacuum polarization. We use the highly-improved staggered quark (HISQ) formulation for the current density with gauge configurations generated with four flavors of HISQ sea quarks. The computation is performed by stochastic estimation of the current density using the truncated solver method combined with deflation of low-modes. The parameters are tuned to minimize the computational cost for a given target uncertainty in the current-current correlation function. The calculation presented here is carried out on a single gauge-field ensemble of size $32^3\times48$ with an approximate lattice spacing of $\sim0.15$ fm and with physical sea-quark masses. We describe the methodology and the analysis procedure.}

\FullConference{The 36th Annual International Symposium on Lattice Field Theory - LATTICE2018\\
		22-28 July, 2018\\
		Michigan State University, East Lansing, Michigan, USA.}

\begin{document}
\section{Introduction}
\label{Sec:Intro}
The anomalous magnetic moment of the muon, defined as
\begin{equation*}
a_\mu = \frac{g_\mu-2}{2},
\end{equation*}
is measured with great precision.  It can also be computed from the Standard Model very precisely.  At present, there is discrepancy of about $3\sigma$ between the theoretical and experimental determination of the quantity.  The current status of the computation, summarized in Ref.~\cite{Giusti:2018mdh}, is
\begin{align*}
a_\mu^{\text{exp}} - a_\mu^{\text{SM}} 
& = 31.3 (4.9)_\text{thy}(6.3)_{\text{exp}} [7.7] \times 10^{-10} \text{ \cite{Jegerlehner:2017lbd}}\\ 
& = 26.8 (4.3)_\text{thy}(6.3)_{\text{exp}} [7.6] \times 10^{-10} \text{ \cite{Davier:2017zfy}}\\ 
& = 27.1 (3.6)_\text{thy}(6.3)_{\text{exp}} [7.3] \times10^{-10} \text{ \cite{Keshavarzi:2018mgv}.}
\end{align*}
The FNAL E989 \cite{Carey:2009zzb} experiment, which recently started its first run, aims to reduce the experimental uncertainty by a factor of four. The goal of the planned J-PARC E34 \cite{JPARC2009p} experiment is to provide a completely independent measurement also with improved precision compared with the Brookhaven experiment.  So a comparable reduction in the theoretical uncertainty is desirable in order to sharpen or resolve the tension.  The largest source of theory error is due to the leading-order hadronic vacuum-polarization (LO HVP) contribution.   In this paper, we focus on the quark-line disconnected part of the HVP contribution to the anomalous magnetic moment of the muon, which complements our ongoing calculation of the leading-order connected HVP contribution.

We describe the first steps in our computation of the contribution to $a_\mu$ from the disconnected part of the hadronic vacuum polarization (HVP).
Other groups have computed this quantity and estimated its value using lattice QCD:
\begin{align*}
a_\mu^{\text{disc HVP}}
&= -11.0 (1.1)_\text{sys}(0.6)_\text{sta} \text{ \cite{Miura18p,Borsanyi:2017zdw}}\\
&= -11.2 (3.3)_\text{sys}(0.4)_\text{F.V.}(2.3)_\text{L}\text{ \cite{Blum:2018mom},}
\end{align*} 
where F.V. indicates a finite volume error and L, long-distance-error \cite{Blum:2015you}.  To further reduce the error and cross-check the various lattice results, we use the highly-improved staggered-quark (HISQ) formulation for the current density with gauge configurations generated with four flavors ($2+1+1$) of HISQ sea quarks.  Here we present our methodology and a preliminary result from one lattice spacing, namely $a\sim0.15$ fm.

\section{Methodology}
\begin{figure}[hb]
\centering
\begin{subfigure}[b]{.5\textwidth}
  \centering
  \includegraphics[width=.4\linewidth]{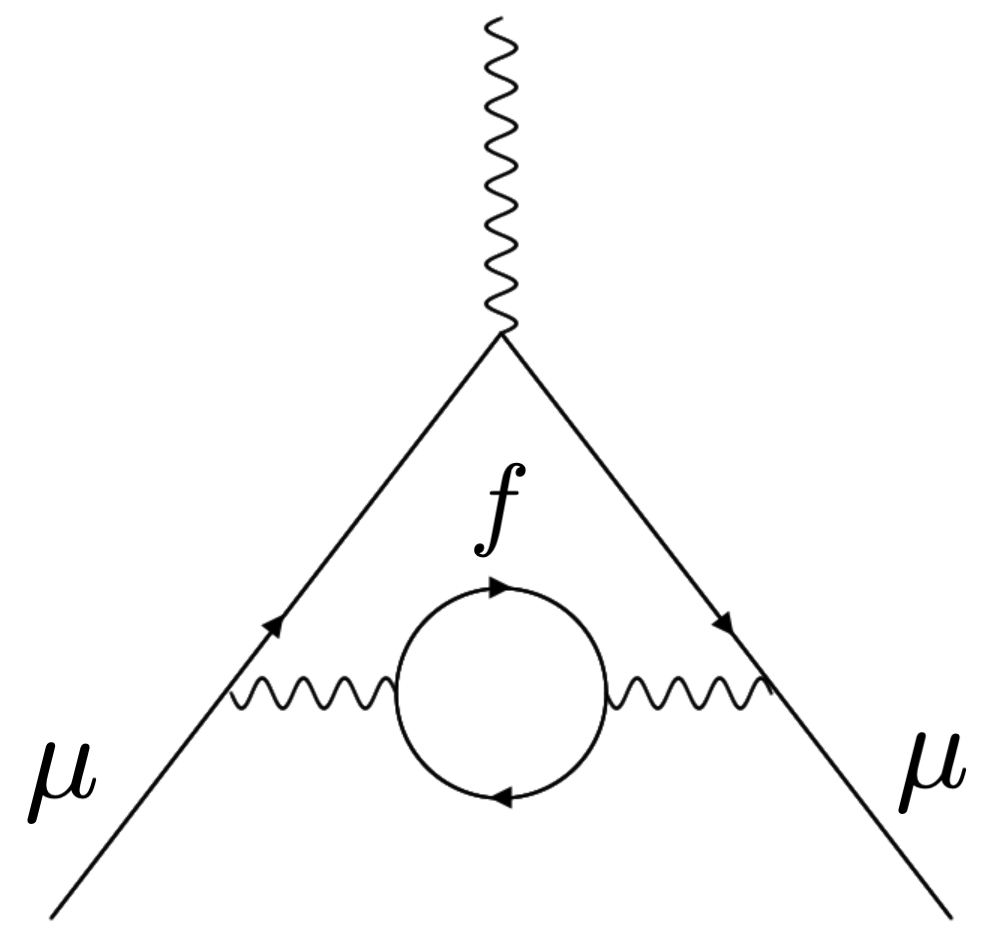}
  \caption{Connected Contribution}
  \label{Fig:ConnDiag}
\end{subfigure}%
\begin{subfigure}[b]{.5\textwidth}
  \centering
  \includegraphics[width=.4\linewidth]{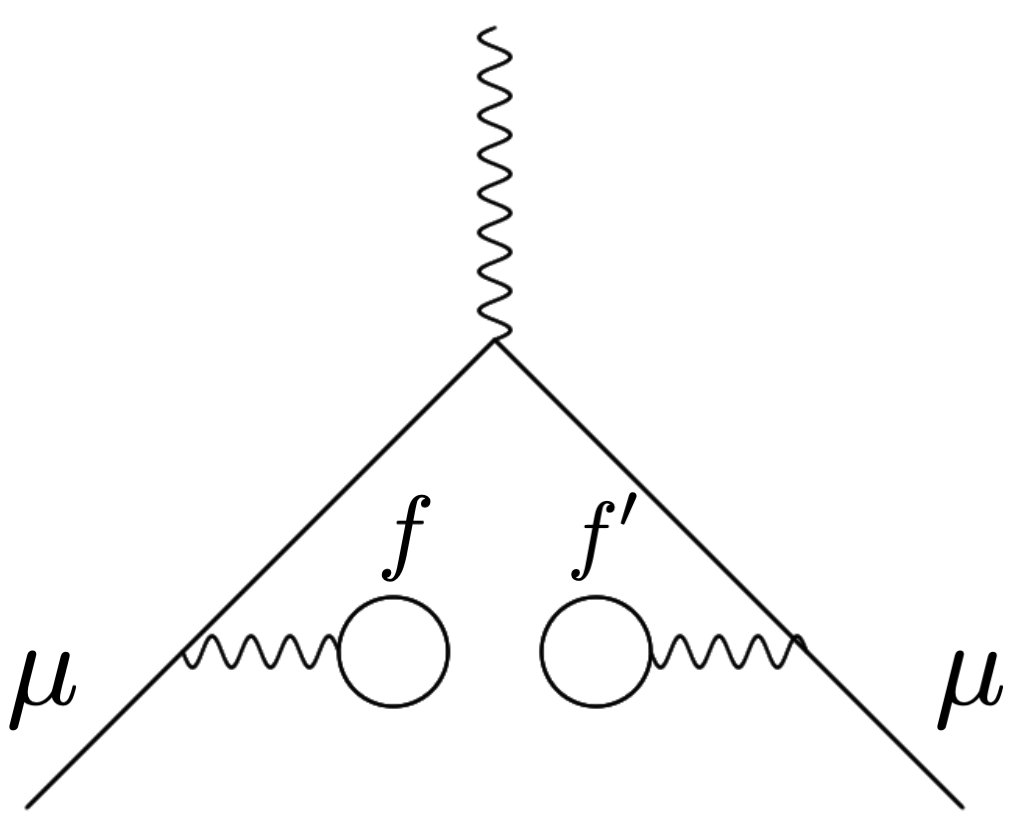}
  \caption{Disconnected Contribution}
  \label{Fig:DiscDiag}
\end{subfigure}
\caption{The Feynman diagrams for the contribution to $a_{\mu}$ from the leading-order HVP.  The quark loop in (\subref{Fig:ConnDiag}) is radiatively corrected by virtual gluons and sea quarks (not shown).  In the same way, the quark loops of flavor $f$ and $f'$ in (\subref{Fig:DiscDiag}) are connected by virtual gluons and sea quarks (not shown).}
\label{Fig:HVPDiag}
\end{figure}

In the continuum, the HVP is given by
\begin{equation}
\Pi^{\mu\nu}(q^2) = (\delta^{\mu\nu} q^2-q^\mu q^\nu)\Pi(q^2)=\int d^4 x e^{iqx}\Braket{J^\mu(x)J^\nu(0)}
\end{equation}
The leading-order contribution  to the anomalous magnetic moment from the HVP \cite{Blum:2002ii} is 
\begin{equation*}
a_\mu^{\text{HVP}} =4 \alpha^2\int_0^\infty dq^2 f(q^2) \hat{\Pi}(q^2).
\end{equation*}
Here, $f(q^2)$ is the kernel function,
\begin{equation*}
f(q^2) =  \frac{m_\mu^2q^2A^3(1-q^2A) }{1+m_\mu^2q^2A^2},
\label{Eq:kernel}
\end{equation*}
where $m_\mu$ is the mass of the muon, $\alpha$ is the fine-structure constant,
\begin{equation*}
A = \frac{\sqrt{q^4+4m_\mu^2q^2}-q^2}{2m_\mu^2q^2},
\end{equation*} 
and $\hat{\Pi}(q^2) = \Pi(q^2)-\Pi(0)$ is the renormalized photon vacuum polarization.  Thus, in the continuum, $\hat{\Pi}(q^2)$ at zero spatial momentum can be computed via \cite{Bernecker:2011gh}
\begin{equation*}
\hat{\Pi}(q^2) = \int_0^\infty dt \left(t^2-\frac{4\sin^2(qt/2)}{q^2}\right)C(t),
\end{equation*}
where $C(t)$ is the time-slice correlation function of the quark-line electromagnetic current.  On the lattice, $C(t)$ is given by
\begin{equation}
C(t) = \frac{1}{3}\sum_{\vec{x},k}\Braket{J_k(\vec{x},t)J_k(0)}
\end{equation}
with $k=1,2,3$ and $J_k = i\sum_f Q_f \bar{\psi}_f\gamma_k\psi_f$ with $f=u,d,s,c$.  Here, $Q_f$ is the charge of the quark of flavor $f$ in units of the electron charge $e$.  The correlator can be divided into two parts: connected and disconnected.  Fig.~\ref{Fig:HVPDiag} shows the schematic diagrams associated with these two pieces. 

In the staggered rooted fermion formulation of lattice fermions, the disconnected part of the HVP correlator is
\begin{equation}
C_{\text{disc}}(t) = \frac{1}{48}\sum_{i=1}^{3}\sum_{\vec{x}}Z_V^2\langle\langle J^{\text{stg}}_i(t,\vec{x})\rangle_F\langle J^{\text{stg}}_i(0)\rangle_F\rangle_G
\label{Eq:discCorr}
\end{equation} 
The additional rooting factor of $1/16$ arises when taking the functional derivative twice with respect to the vector potential to obtain Eq.~(\ref{Eq:discCorr}).
The disconnected part of the HVP can then be obtained from
\begin{equation*}
\Pi^{\text{disc HVP}}(q^2) = a^4\sum_{t}e^{iqt}C_{\text{disc}}(t)
\end{equation*}

In Eq.~(\ref{Eq:discCorr}), $F$ indicates the integration over the fermionic degrees of freedom, which is performed explicitly, whereas averaging over gauge configurations is indicated by $G$.  $Z_V$ is the vector-current renormalization factor.  $J^{\text{stg}}_i(r)$  is the one-link current with $\Gamma_\mu\otimes\Gamma_t = \gamma_\mu\otimes\mathbf{1}$.

We use stochastic estimation of the current density using Z(2) random sources with support on the entire four-volume $\Omega$.  We use the truncated solver method (TSM) \cite{Alexandrou:2012zz, Bali:2009hu} combined with low-mode deflation.  Dilution with stride $2$ is also applied to the random sources to reduce the variance \cite{Endress:2014qpa}.  That is, on each lattice, we use 16 random sources, each with support on a subset $\Omega_\eta = \{r\in\Omega\,|\,\eta \equiv r \text{ (mod }2)\}$ where $\eta_\mu=0$ or $1$.
The low-mode part is computed exactly by constructing it from the eigenvectors of $\slashed{D}$.  

Then, with $J^\text{stg}_i(r) = \sum_{f}Q_f j_{f,i}(r)$, the current density is computed via $j_{f,i}(r)$, which is given as follows:
\begin{equation}
j_{f,\mu}(r)=j^{(p)}_{\text{low},f,\mu}(r)+j_{f,\mu}^{(\text{sloppy})}(r) +j_{f,\mu}^{(\text{diff})}(r),
\label{Eq:jmu}
\end{equation}
where $j^{(p)}_{\text{low},f,\mu}(r)$ is the low-mode part of the current constructed from the eigenpairs, and the rest is the stochastically estimated high-mode part.  If we let $\xi$ represent the random source, without deflation the stochastic estimation of $j_{f,\mu}(r)$ is given by
\begin{equation*}
j_{f,\mu}(r) = \frac{1}{N_r}\sum_{k=1}^{N_r} \text{Im Tr}_c\left[\xi^\dagger_k(r)\alpha_\mu(r)U_\mu(r) (M_f^{-1}\xi_r)(r+\hat{\mu})\right]
\end{equation*}
where $\alpha_\mu(r)=(-1)^{\sum_{\nu<\mu}x_\nu}$ are the usual staggered phases, $U_\mu(r)$ the gauge link, and $M_f$ is the fermion matrix for the flavor $f$.  The trace is taken over colors.  In Eq.~(\ref{Eq:jmu}), $j^{\text{(sloppy)}}_{f,\mu}=j_{f,\mu}^{(\text{high})}\big|_{(\text{sloppy})}^{N_r=N_{\text{sloppy}}} (r)$ is the stochastically estimated high-mode part with sloppy inversion with $N_\text{sloppy}$ random sources, and $j^{\text{(diff)}}_{f,\mu}(r)$, the difference in the stochastic estimates with $N_\text{fine}$ sloppy and fine inversions, i.e., 
\begin{equation*}
j^{\text{(diff)}}_{f,\mu}(r) =j_{f,\mu}^{(\text{high})}\big|_{(\text{fine})}^{N_r=N_{\text{fine}}}(r) -j_{f,\mu}^{(\text{high})}\big|_{(\text{sloppy})}^{N_r=N_{\text{fine}}} (r).
\end{equation*}


\section{Parameter Optimization}

The calculation presented here is carried out on a single gauge-field ensemble of size $32^3 \times 48$ with an approximate lattice spacing of $0.15$ fm using TSM and deflation along with dilution with stride $2$.  Accordingly, there are several parameters in this simulation that need to be tuned to achieve the target statistical uncertainty due to stochastic estimation in the current-current correlation function at minimum computational cost.  The final statistical error coming from gauge fluctuations is reduced by increasing the number of gauge configurations analyzed in the tuned parameter setting.  

The tuning starts by setting a goal uncertainty in the statistical estimate of the correlator for a single configuration.  We have set this to be $1\%$ of the average value of the correlator.  Then, we go on to determine the precision of the fine solve required to ensure that the systematic uncertainty due to the finite solver precision is less than the 1\% accuracy.  The residual of the sloppy solve and the ratio of the number of sloppy solves to fine solves are determined to minimize the uncertainty at a given computational cost.  The remaining parameter of the simulation, namely the number of sloppy solves and deflating eigenpairs are tuned so as to give the desired precision in the correlator at a minimal computational cost \cite{Alexandrou:2013wca}.  The residuals of the eigenpairs $\big| D_{eo}D_{eo} \tilde{v}_n^{(e)}-\tilde{\lambda}_n^2\tilde{v}_n^{(e)} \big| $ are set so that the first few smallest estimated eigenvalues are closer to their true eigenvalues where $\tilde{v}_n^{(e)}$ and $\tilde{\lambda}_n$ are the estimated $n^{\text{th}}$ eigenpair.  They were set to $10^{-9}$.   After tuning, the optimum parameter values are found to be
\begin{itemize}
\item The number of eigenpairs for deflation: $350$
\item The residual of fine and sloppy solve: $2.70\times10^{-2}$ and $1\times10^{-5}$, respectively
\item The number of fine and sloppy solves per configuration: $72$ and $1408$, respectively. 
\end{itemize}
In order for deflation to be effective, the residual of the eigensolutions should be smaller than the desired residual of the sparse-matrix solution \cite{Davies:2017mzz}.  So the residual can be set higher than $10^{-9}$ without affecting performance of the inversion.  This residual is not tuned in this work.  

\section{Result}
In this preliminary calculation, $C_{\text{disc}}(t)$ is computed using $1326$ gauge configurations with physical sea-quark masses.  Fig.~\ref{Fig:timeCorr} shows our result for $C_{\text{disc}}(t)$ at a lattice spacing of $\sim0.15$ fm.  In the plot, an opposite-parity oscillating component is clearly visible, as expected with staggered fermions.

\section{Analysis}
\begin{figure}
\centering
\begin{minipage}{.49\textwidth}
  \centering
  \includegraphics[width=1.1\linewidth]{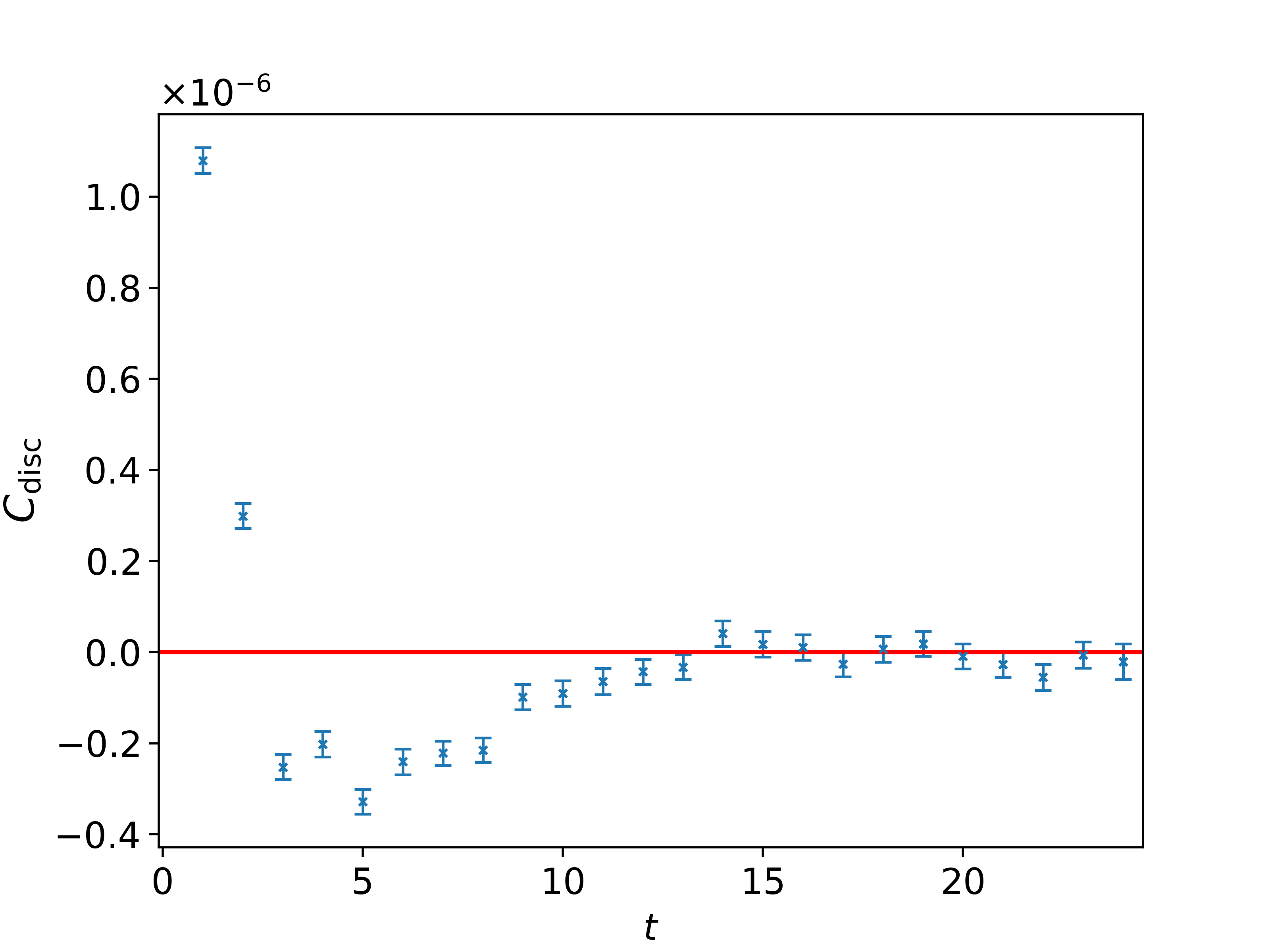}
  \captionof{figure}{Time-slice disconnected current density correlator vs. the temporal separation in lattice units.}
  \label{Fig:timeCorr}
\end{minipage}%
\hfill
\begin{minipage}{.49\textwidth}
  \centering
  \includegraphics[width=1.08\linewidth]{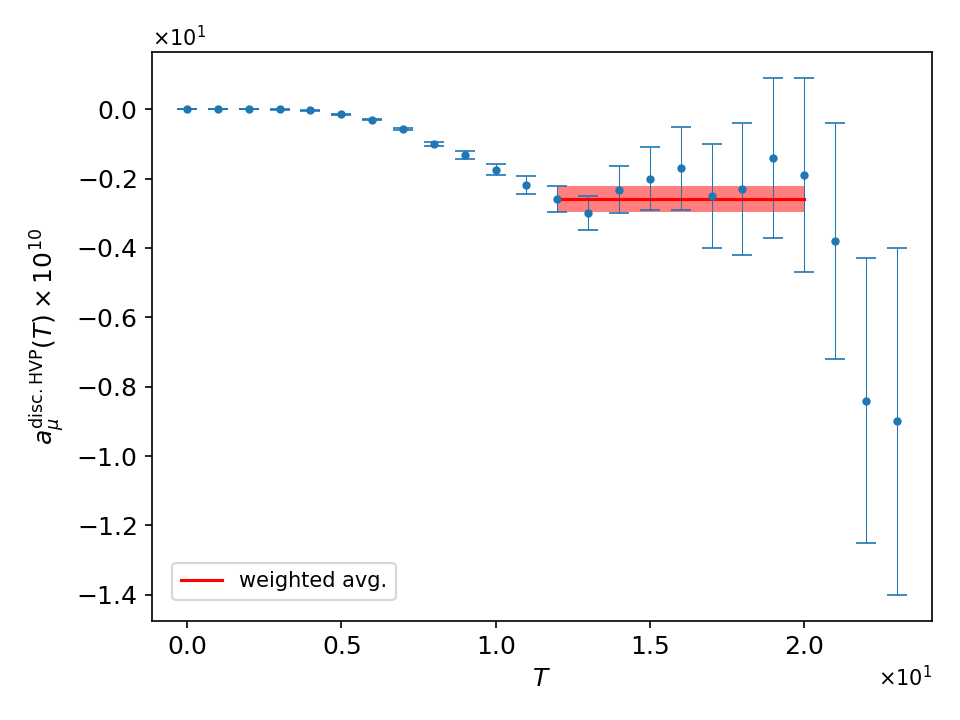}
  \captionof{figure}{$a_\mu^{\text{disc}}\times10^{10}$ as a function of the cut-off time $T$ in the TMR in lattice units. A constant fit is taken from $T=12$ to $20$.  The shaded region indicates the error.}
  \label{Fig:amu_T}
\end{minipage}
\end{figure}

To compute the disconnected contribution to $a^{\text{HVP}}_\mu$, we use the time-moment representation (TMR) \cite{Blum:2015you}, i.e.,
\begin{equation*}
a^{\text{disc. HVP}}_\mu = \sum_{t=0}^{T} w_t C_{\text{disc}}(t) 
\end{equation*}
where
\begin{equation*}
w_t = 4\alpha^2\int_{0}^{\infty}dq^2 f(q^2)\left( t^2-\frac{4\sin^2(qt/2)}{q^2}\right).
\end{equation*} 
With the value of $Z_V^{-1} = 0.837(3)$ \cite{Koponen:2015tkr}, the values of $a_\mu^{\text{disc}}$ as a function of $T$ in lattice units is shown in Fig.~\ref{Fig:amu_T}.  By choosing the fit range of $(12,20)$, we obtain a preliminary weighted average of 
\begin{equation}
a_\mu^{\text{disc}} = -2.5(5)\times10^{-10}
\end{equation}
This value is less negative than the value cited in Sec.~\ref{Sec:Intro}.  This is expected.  As is observed in Ref.~\cite{Borsanyi:2017zdw}, $a_\mu^{\text{disc}}$ has a strong lattice-spacing dependence; the smaller the lattice spacing is, the more negative $a_\mu^{\text{disc}}$ becomes.


\section{Conclusion}

We have computed the disconnected part of the anomalous magnetic moment of the muon on a lattice of size $32^3\times48$ with the lattice spacing of $a\sim0.15$fm.  The value we obtained was $-2.5(5)\times10^{-10}$.  For the future, we need to increase the statistics and compute $a_\mu^{\text{disc}}$ at multiple lattice spacings.

\section{Acknowledgments}
Computation for this work was done using resources of the National Energy Research Scientific Computing Center (NERSC), a U.S. Department of Energy Office of Science User Facility operated under Contract No. DE-AC02-05CH11231.  S.Y., C.D., and A.V. are supported by the U.S. National Science Foundation under grants PHY14-14614 and PHY17-19626. C.M. is supported by the STFC Consolidated Grant  ST/P000479/1.  R.V. is supported by the Fermi National Accelerator Laboratory (Fermilab), a U.S. Department of Energy, Office of Science, HEP User Facility.  Fermilab is managed by Fermi Research Alliance, LLC (FRA), acting under Contract No. DE-AC02-07CH11359. A.X.K. is supported by No. DE-SC0015655.

\bibliographystyle{JHEP}
\bibliography{Lattice2018Proceeding_v2}

\end{document}